\documentclass[11pt]{article}
\usepackage{sectsty}
\usepackage{authblk}
\usepackage{graphicx}
\usepackage{amsmath}

\usepackage{qcircuit}
\usepackage{fancyhdr}
\usepackage{ifoddpage}
\usepackage{blindtext}
\usepackage{amssymb}

\usepackage{todonotes}

\newtheorem{claim}{Claim}

\newcommand{\myCirc}[1]{\mbox{\Qcircuit @C=1.em @R=1.2em {#1}\null\mbox{}} }

\def\mytitle{Machine Learning Kernel Method from a Quantum Generative Model}
\def\mytitlefull{Machine Learning Kernel Method	from a Quantum Generative Model }

\def\myauthor{Przemys{\l}aw Sadowski}

\title{\sffamily\bfseries \mytitlefull}
\author{\myauthor\thanks{psadowski@iitis.pl}}
\affil{\small \textit{Institute of Theoretical and Applied Informatics,} \textit{Polish Academy of Sciences,}\\\textit{Ba{\l}tycka 5, 44-100 Gliwice, Poland}}

\let\Sectionmark\sectionmark
\def\sectionmark#1{\def\Sectionname{#1}\Sectionmark{#1}}
\let\Subsectionmark\subsectionmark
\def\subsectionmark#1{\def\Subsectionname{#1}\Subsectionmark{#1}}
\let\Subsubsectionmark\subsubsectionmark
\def\subsubsectionmark#1{\def\Subsubsectionname{#1}\Subsubsectionmark{#1}}

\def\randomvec{\mathbf{u}_i}
\def\randomvecg{\mathbf{g}_i}
\def\inputvec{\mathbf{f}_j}
\def\normalized{\mathbf{d}_j}
\def\outputvec{\mathbf{s}_{i,j}}
\def\realampli{a_{i,j}}
\def\multiplied{b_{i,j}}
\def\resultc{c_{i,j}}
\def\results{c_{D+i,j}}
\def\weight{w_{i}}
\def\Complex{\mathbb{C}}
\def\Real{\mathbb{R}}

\newcommand{\inner}[2]{\langle #1 , #2 \rangle}
\newcommand{\real}[0]{\mathbb{R}}

\begin{document}
\allsectionsfont{\normalfont\sffamily\bfseries}

\maketitle
\thispagestyle{empty}
\renewcommand{\footrulewidth}{0.4pt}
\begin{abstract}
Recently the use of Noisy Intermediate Scale Quantum (NISQ) devices for machine learning tasks has been proposed.
The propositions often perform poorly due to various restrictions.
However, the quantum devices should perform well in sampling tasks.
Thus, we recall theory of sampling-based approach to machine learning and 
propose a quantum sampling based classifier.
Namely, we use randomized feature map approach.
We propose a method of quantum sampling based on random quantum circuits with parametrized rotations distribution.
We obtain simple to use method with intuitive hyper-parameters that performs at
least equally well as top out-of-the-box classical methods.
In short we obtain a competitive quantum classifier with crucial component
being quantum sampling -- a promising task for quantum supremacy.
\end{abstract}

\section*{Introduction}
The use of Noisy Intermediate Scale Quantum (NISQ) devices for machine learning tasks has been proposed in various
forms~\cite{biamonte2017quantum}.
However, there is still a lot to be understood
about the potential sources of quantum advantage.
While it is reasonable 
to postulate that quantum computers may outperform classical computers
in machine learning tasks,
the existing propositions often perform poorly due to various restrictions~\cite{ciliberto2018quantum}.

As some people put it, NISQ devices are best at simulating themself.
In particular, this means performing noisy random quantum circuits.
It may seem to be not a very useful task, however
the noisy devices has been successfully used
as a resource 
for so called quantum generative models,
leading to non-trivial probability distributions in vector spaces~\cite{preskill2018quantum, aaronson2016complexity, boixo2018characterizing, romero2019variational}.
Thus, maybe we could harness that specific resource for machine learning tasks.

At the same time, is has been shown that any translation invariant kernel
could be substituted by a explicit feature map based on some probability distribution
in the vector space of the features~\cite{rahimi2008random}.
Putting it the other way around, any unique probability distribution
on a vector space of features
can be expected to
raise a new kernel for that features space.

We recall the theory of sampling-based approach to machine learning
and
establish a link to quantum generative models.
Our goal is to show that we could develop quantum machine learning methods
that use quantum devices solely as a source of a probability distribution, hoping to efficiently use
the quantum resource.

We use a scheme designed for random features for large-scale kernel machine~\cite{rahimi2008random}
as sampling based classification.
We propose a method of
quantum sampling based on random quantum circuits with parameterized rotations
distribution.
In short we obtain a competitive quantum classifier with crucial component
being quantum sampling -- a promising task for quantum supremacy.

\section{Preliminaries -- Randomized Feature Maps}\label{sec:original-scheme}
\begin{figure}
\centering
\includegraphics[]{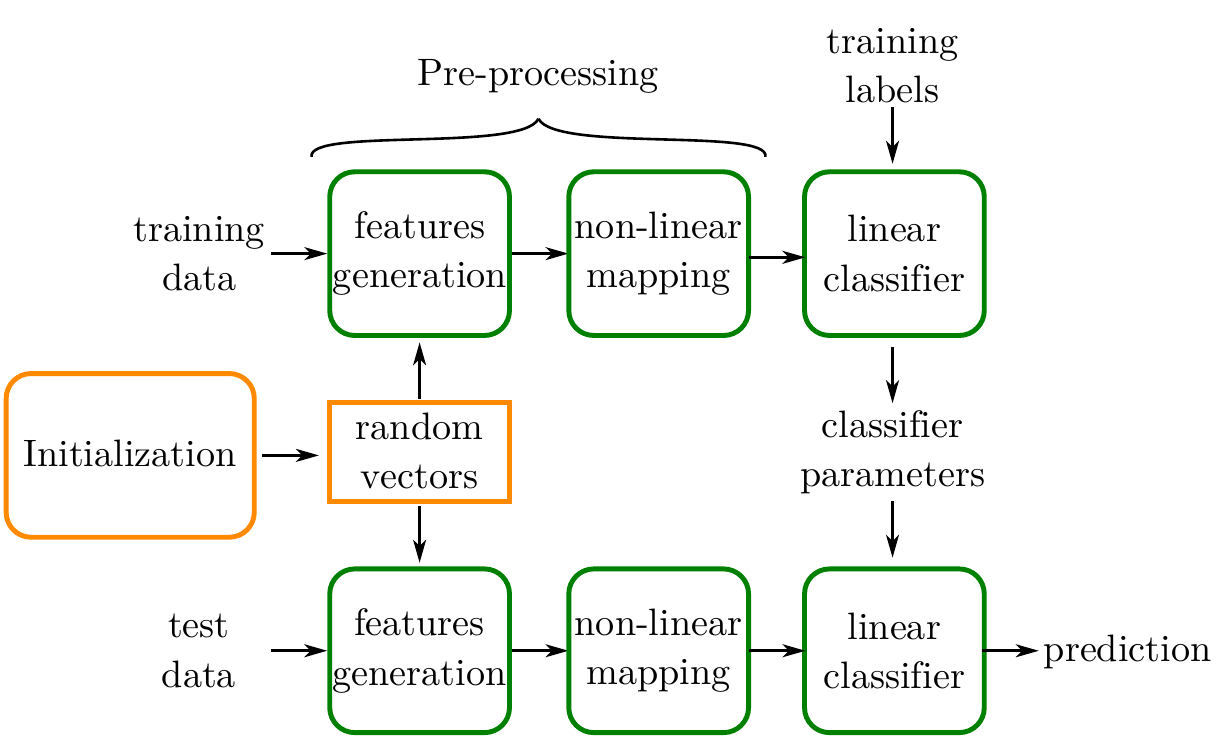}
\caption{Original random features generation scheme. The crucial part that
	decides what kernel is effectively implemented is random vectors
	sampling, \textit{i.e.} initialization.}\label{fig:sinks}
\end{figure}

One method to tackle large-scale kernels has been proposed by Rahimi and
Recht~\cite{rahimi2008random, rahimi2009weighted}.
The proposition is to do pre-processing, mapping the input data
to a randomized low-dimensional feature space
and then apply a linear
classifier.
The explicit pipeline is shown in Figure~\ref{fig:sinks}.
The key of the idea is to replace designing a fancy kernel with developing a sophisticated method for sampling vectors.
We will introduce the original idea, split it into steps and specify which ones will be further considered.

Let us recall one of the key theorems that provides a foundation for the proposed scheme. 
Lets assume we have an 
input feature space $\mathcal{X}=\real^d$ and
a shift invariant kernel $k:\mathcal{X}\times \mathcal{X}\rightarrow \real$,
$P$ is the corresponding probability distribution in the input features space $\mathcal{X}$, 
$c:\mathcal{X}\rightarrow\mathcal{Y}=\real^{2D}$ is an explicit map into a higher dimensional space built in a certain way based on 
random variable $g$ sampled with distribution $P$.
\begin{claim}(Uniform convergence of Fourier features)~\cite{rahimi2008random}.
Let $M$ be a compact subset of $\real^d$ with diameter $diam(M)$. Then, for the mapping $c$, we have
\begin{equation}
Pr[\sup_{\mathbf{f_1},\mathbf{f_2}\in M}|c(\mathbf{f_1})^\intercal c(\mathbf{f_2})-k(\mathbf{f_1},\mathbf{f_2})|\ge \epsilon]\le
2^8\left(\frac{\sigma_P diam(M)}{\epsilon}\right)^2\exp(-\frac{D\epsilon^2}{4(d+2)}),
\end{equation}
where $\sigma_P^2 = E_P[g^\intercal g]$ is the second moment of the Fourier transform of $k$.
Further, $\sup_{\mathbf{f_1},\mathbf{f_2}\in M} |c(\mathbf{f_1})^\intercal c(\mathbf{f_2})-k(\mathbf{f_1},\mathbf{f_2})|\le \epsilon$
with any constant probability when
$D =\Omega(\frac{d}{\epsilon^2}\log\frac{\sigma_Pdiam(M)}{\epsilon})$.
\end{claim}

The proposition from~\cite{rahimi2008random} for the construction of $P$ and $c$ was to sample a set of $D$ vectors $\mathbf{g_i}\sim P$, $i=1,\ldots,D$ of the same dimension as
the data at hand as a basis for the new features. The vectors in the training data are
compared with the random set and new features are generated as
of the corresponding vectors. Then all features are transformed
separately
via a one dimensional non-linear (cosine, sine)
function
\begin{equation}
c(\mathbf{f}) = \sqrt{\frac{1}{D}}[ \cos(\mathbf{g_1}^\intercal\mathbf{f}), \ldots, \cos(\mathbf{g_D}^\intercal\mathbf{f}), \sin(\mathbf{g_i}^\intercal\mathbf{f}), \ldots, \sin(\mathbf{g_D}^\intercal\mathbf{f})]^\intercal
\end{equation}
 New data set is passed on to a linear classifier for training.
New test data is transformed the same way. The inner product is computed
using the same random vectors sample and transformed with the same non-linear
function. Then, the previously trained linear classifier is used.

The scheme can be seen as a typical classifier with pre-processing phase.
In such picture we have the following steps:
\begin{itemize}
	\item Initialization.
	\item Pre-processing of the training data.
	\item Classifier training with the processed data.
	\item Pre-processing of the test data.
	\item Applying the classifier on the processed test data.
\end{itemize}
To implement the scheme we need to specify the initialization and pre-processing steps.

We split the pre-processing into features generation and non-linear map and will focus mostly on the former one later in the paper.
We operate on data points with dimension $d$ and create $D$ new features that does not have to be equal $d$.
\begin{itemize}
	\item Initialization: Sample $\mathbf{g_i}\in \real^d$, $i=1,\ldots, D$ from a distribution given by hyper-parameters.
	\item Pre-processing: Transform any given training or test vector $\mathbf{f}\in\real^n$ into
	$\mathbf{c}=( \cos(\inner{\mathbf{f}}{\mathbf{g_1}}), \ldots, \cos(\inner{\mathbf{f}}{\mathbf{g_{D}}}), \sin(\inner{\mathbf{f}}{\mathbf{g_{1}}}),\ldots, \sin(\inner{\mathbf{f}}{\mathbf{g_{D}}}) )\in\real^{2D}$.
\end{itemize}
In the remaining part we will show how to implement these steps using quantum circuits for vectors sampling.

\section{Features Generation from Quantum Circuits}\label{sec:features-from-circuits}

For our purposes it should be sufficient to know that
any program that can be run on a $k$-qubit quantum computing
device can usually be described by a unitary operator $U\in
\mathcal{U}(\Complex^{2^k})$.
In this notation for an input vector $\mathbf{d}\in\Complex^{2^k}$, the output
$\mathbf{s}\in\Complex^{2^k}$ is simply a result of multiplication
\begin{equation}
\mathbf{s} = U \cdot\mathbf{d}.
\end{equation}
In this context the considered vectors $\mathbf{s}, \mathbf{d}$ are normalized and called states.

\begin{figure}
	\centering
\includegraphics[]{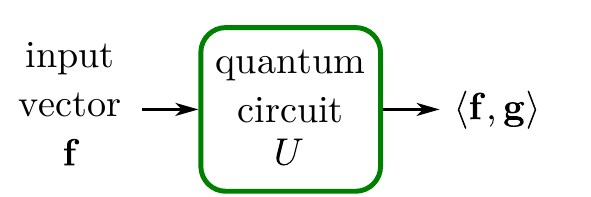}
\caption{Quantum part of the scheme in the most general picture.
We link vector $\mathbf{g}$ with quantum operation $U$, such that the result of the operation for any input vector $\mathbf{f}$
provides information about the inner product $\inner{\mathbf{f}}{\mathbf{g}}$.}\label{fig:basic}
\end{figure}

We aim at linking random vectors to quantum circuits in such a way that allows us to compute the inner product.
Such scheme would be compatible with the scheme in the previous section.
The quantum part in the most general picture is presented in Fig.~\ref{fig:basic}.

In particular, we fix a quantum operation $U$ and denote the $1^{\mathrm{st}}$ row (Hermitian conjugated) as
\begin{equation}
\mathbf{u}=U^\dagger\cdot\mathbf{z},
\end{equation}
 using basis vector $\mathbf{z}=(1,0,\ldots,0)$.
For  given $\mathbf{d}$ we want to compute the inner product $\inner{\mathbf{u}}{\mathbf{d}}$.
Let us note that we have
\begin{equation}
	\inner{\mathbf{z}}{\mathbf{s}}=s_1,
\end{equation}
when $\mathbf{s}=(s_1, s_2, \ldots, s_{2^k})$.
We can obtain inner product with some $\mathbf{d}$ on a quantum computer by injecting $\mathbf{d}$ as the input state and
reading the $1^{\mathrm{st}}$ value of the output state
\begin{equation}
s_1 = (U\cdot \mathbf{d})_1 = \inner{\mathbf{z}}{U\cdot\mathbf{d}} = \inner{U^\dagger\cdot\mathbf{z}}{\mathbf{d}}=\inner{\mathbf{u}}{\mathbf{d}}.
\end{equation}
Let us note that reading the exact value requires so called \textit{state
tomography}~\cite{thew2002qudit} and can be done with arbitrary
precision with arbitrarily high probability, but always approximately.
In practice esimating this value is a Bernoulli trial.

Based on the above considerations the randomized feature map is constructed as
follows.
Based on some quantum operation construction procedure we define a probability distribution
on a set of unitary operators. This set will be described in detail later as
circuit Ans\"atze. The parameters of this distribution are hyper-parameters of
the whole classification scheme.
From fixed distribution we sample a set of unitary operators $\{U_i\}$,
and in consequence vectors $\mathbf{u_i}$.
Each vector $\mathbf{u_i}$ is a row of a unitary operator and thus normalized.
To obtain vectors of variable length we sample the lengths $\weight\in\Real$.
The resulting set of vectors is $\randomvecg = w_i\mathrm{Re}(\randomvec)$,
where
$\randomvec=U_i^\dagger\cdot\mathbf{z}$,
for $i=1,\ldots, D$.
Let us note that the length $\weight$ can be correlated with the sampled vector.
For example, we could design a larger circuit, with additional qubits, that after the
measurement would indicate the length.

\begin{figure}[t]
\centering
\includegraphics[]{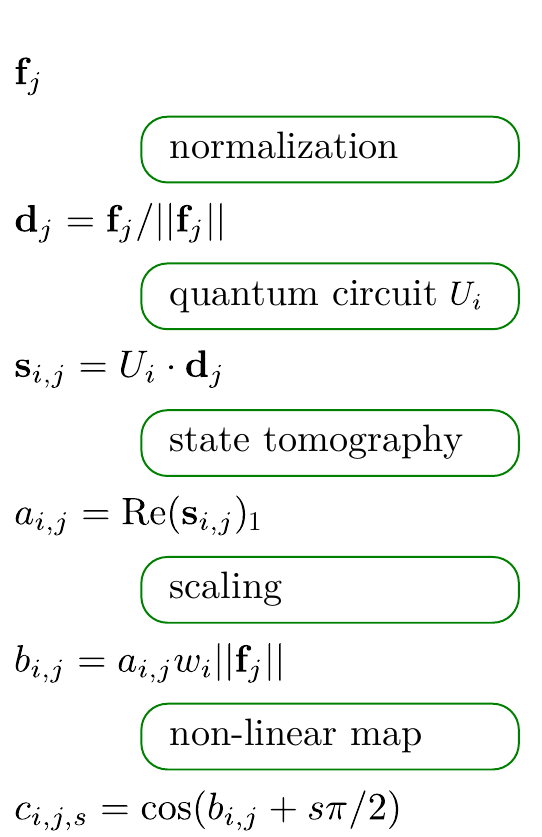}
\caption{Random features calculation with quantum circuits. For an input feature
	vector $\mathbf{f}_i$, a feature $c_{i,j}$ is constructed using a quantum
circuit unitary operation $U_i$ with weight $w_i$.
for all random circuits.}\label{fig:steps}
\end{figure}

The mapping is performed as follows.
In order to obtain feature $i,s$ for data point $j$ we 
\begin{itemize}
	\item map a data point $\inputvec$ into a normalized vector $\normalized$ and store $||\inputvec||$,
	\item apply circuit $U_i$ to state $\normalized$, computing $\outputvec=U_i\cdot\normalized$,
	\item estimate real part of the $1^{\mathrm{st}}$ amplitude of $\outputvec$, obtaining
	 $\realampli=\mathrm{Re}(\outputvec)_1=\mathrm{Re}(\inner{{\randomvec}}{{\normalized}})$,
	\item scale the vectors obtaining $\multiplied=\realampli\weight||\inputvec||=\inner{\randomvecg }{\inputvec}$,
	\item apply a non-linear map cos/sin obtaining $\resultc = \cos( \multiplied)$, $\results = \sin( \multiplied)$,
	\item return $\resultc, \results$.
\end{itemize}
Effectively $C=\{c_{i,j}\}$ is a concatenation of $\cos( G\cdot F)$ and $\sin( G\cdot F)$, where cos/sin are element-wise matrix
operations,
rows of $G$ are random vectors $\randomvecg$ and columns of $F$ are
data points $\inputvec$.
The steps are sketched in Fig.~\ref{fig:steps}.

\section{Example}

\subsection{Quantum Circuits}
In the previous section we stated that we use quantum operations to generate the features.
In Section~\ref{sec:features-from-circuits} we only mentioned that any operation
corresponds to an unitary operation.
However, we will use a much more practical way of defining quantum operations --
quantum circuits. That is a computational model inspired by classical logic
circuits. The common idea is to describe a complex global operation with a
sequence of simple and small basic operations.
The most often recommended introduction can be found in~\cite{nielsen2010quantum}.

We will use a set of basic operations represented by two parameterized unitary operations
and build circuits by multiplying these matrices.
The operations will correspond to one qubit
rotations $R_y$ and a two-qubit entangling gate CNOT. The matrix representations
of the two are
\begin{equation}
\begin{split}
R_y(\alpha) = e^{-i\sigma_y\alpha}=
\left(\begin{matrix}
 \cos(\alpha) & \sin(\alpha) \\
-\sin(\alpha) & \cos(\alpha)
\end{matrix}\right),\\
\mathrm{CNOT}=\left(\begin{matrix}
1 &0 &0 & 0 \\
0 &1 &0 &   0\\
0 &0 &0 & 1 \\
0 &0 &1 &   0
\end{matrix}\right),
\end{split}
\end{equation}
where $\sigma_y$ is a Pauli matrix.

The end operation $U$ acts on a bigger space than $R_y$ or
CNOT. We will assume that all of the gates are extended to the same space with
tensor product operation.
Thus we will specify on which subspace the operator works.
In case of $R_y(\alpha)$ the operator works on a subspace corresponding to one qubit and
we will use
\begin{equation}
R_y^{(j)}(\alpha) = 1_{2^{j-1}} \otimes R_y(\alpha) \otimes 1_{2^{k-j}}
\end{equation}
to mark that it is the $j$th out of $k$ qubits,
where $1_{d}$ is a $d$-dimensional identity matrix and $\otimes$ is tensor product operation (also tensordot in \textit{e.g.} numpy). In case of
CNOT
the operator works on a product of two subspaces, corresponding to so called
\emph{target} and \emph{control} qubits. We will mark it with CNOT$^{(i,j)}$.

\subsection{Random Vectors Circuit Ans{\"a}tze}
For this example we chose an Ans{\"a}tze that generate a
broad family of quantum circuits with little hyper-parameters that
have intuitive interpretation.
The parameters that will need to be fixed are the number of layers $L$, the parameters of
the normal distribution used for rotations: $m, \sigma$, and variance of the vectors length $\sigma_w$.

\begin{figure}[h]
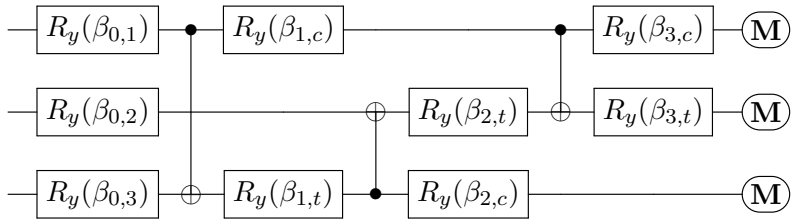

	\begin{center}
		\mbox{
			\myCirc {
				& \gate{R_y(\beta_{0,1})} & \ctrl{2}    & \gate{R_y(\beta_{1,c})}& \qw       &\qw        		 &\ctrl{1}  &\gate{R_y(\beta_{3,c})}&\measure{\mathbf{M}} \\
				& \gate{R_y(\beta_{0,2})} & \qw         & \qw          		 & \qw\oplus &\gate{R_y(\beta_{2,t})}&\qw\oplus       &\gate{R_y(\beta_{3,t})}        	    &\measure{\mathbf{M}}\\
				& \gate{R_y(\beta_{0,3})} & \qw\oplus   & \gate{R_y(\beta_{1,t})}& \ctrl{-1}  &\gate{R_y(\beta_{2,c})}&\qw &\qw&\measure{\mathbf{M}} 
			}
		}
	\end{center} \vspace{1ex}
	\caption{Circuit Ans{\"a}tze example. The circuit always begins with a rotation
	on each of the qubits. Then, a number of CNOTs is put at random qubits, each
	followed with rotations on both target ($\oplus$) and control ($\mathbf{\bullet}$) qubits. All of the rotation
	parameters are sampled from the same distribution described by the
	hyper-parameters. Here
	$U= R_y^{(1)}(\beta_{3,c}) R_y^{(2)}(\beta_{3,t})
		\mathrm{CNOT}^{(1,2)} R_y^{(2)}(\beta_{2,t}) R_y^{(3)}(\beta_{2,c})
		\cdot
		\mathrm{CNOT}^{(3,2)} R_y^{(1)}(\beta_{1,c}) R_y^{(3)}(\beta_{1,t})
		\mathrm{CNOT}^{(1,3)}
		R_y^{(1)}(\beta_{0,1}) R_y^{(2)}(\beta_{0,2})R_y^{(3)}(\beta_{0,3})
		$.}\label{fig:ansatze-example}
\end{figure}

For $k$ qubits the circuit is created as follows. First a rotation gate $R_y$ is added
on each of the qubits, $k$ gates in total.
\begin{equation}
U_0 = \prod_{j=1}^k R_y^{(j)}(\beta_{0,j}).
\end{equation}
Then for layer $l=1,\ldots, L$ we repeat: sample
a control and action qubits $q_{l,c}, q_{l,t}$ for a CNOT gate and then add a rotation on
both action and control qubits
\begin{equation}
U_l = \mathrm{CNOT}^{(q_{l,c}, q_{l,t})} R_y^{(q_{l,c})}(\beta_{l,c}) R_y^{(q_{l,t})}(\beta_{l,t}).
\end{equation}
The resulting operator is composed as
\begin{equation}
U = U_L U_{L-1} \ldots U_1 U_0
\end{equation}
An example is
presented in Figure~\ref{fig:ansatze-example}. The rotation angles are sampled from the distribution
described by the hyper-parameters, $\beta_{l,j}\sim\mathrm{\texttt{gauss}}(m, \sigma)$.
We use Gaussian distribution with fixed
mean and variance.
An additional hyper-parameter is $\sigma_w$ that will affect the weights of the vectors. For circuit $i$ we store
$w_i \sim \mathrm{\texttt{gauss}}(1, \sigma_w)$ as the weight corresponding to circuit $i$, so that we will effectively
consider vector $\mathbf{g_i}=\mathrm{Re}( w_i\randomvec )$.

\subsection{Setting}
In this work we perform basic accuracy measuring experiments.
As the testing dataset we consider the MNIST dataset~\cite{shamsuddin2018exploratory} as in~\cite{wilson2019quantum}.
We aim at beating the SVM with radical basis function as the kernel.
We explore the space of hyper-parameters and the relation between the score and
the number of random quantum circuits used.

The whole experimental algorithm is based on the randomized feature maps scheme
presented in Section~\ref{sec:original-scheme}. We first describe the details of
the preprocessing and classification used, and then report the obtained results.

The MNIST dataset contains 70000 images corresponding to digits 0-9. 
We extract only two of the digits: 3 and 5.
There are 13454 data points of either of them.
For measuring the accuracy we use a single training-test pair with size proportion 6:1.
The size of the
sets are 11532, and 1922 correspondingly.

Before feeding the algorithm with data we do simple feature selection.
We plan to use a 7 qubit circuits that operate on vectors of dimension equal to $2^7=128$.
Thus we select 128 best features according to a $\chi^2$ test, looking for multimodal distributions.

For features selection we use \texttt{SelectKBest} method
{and we perform classification with \texttt{LinearSVC} method from \texttt{sklearn}~\cite{scikit-learn}.}

\subsection{Scores}

In the presented example we analyse the accuracy of the resulting classification
scheme. We will compare the results to the ones obtainable by linear and
non-linear methods. The results depend on the hyper-parameters selection, thus
we show the results obtained for a range of values.

The accuracy considered here is the fraction of correct answers in a binary
classification scheme. 
For comparison we take permutation invariant methods, without any optimisation towards image processing.
The two main reference points are Linear SVC and SVM with radical basis function kernel.
The scores obtainable with these methods are roughly 96\% and 99\%.

One particularly important hyper-parameter is selected circuits size.
This is the one that is connected to the complexity of the quantum part of the
scheme.
Larger circuit size would 
increase both simulation cost and quantum device running time.
In the case of the selected Ans{\"a}tze we can select the number of CNOT gates
freely.
In this example we consider the number of CNOT gates being multiplication of the
number of qubits.

Our hyper-parameter selection has been done with grid search for small number of random vectors $D$.
The best score was equal to 0.9909 for $D=8000$,
although the average sore was highest for the largest tested value of $D=16000$.
This result is better than what we achieved using SVM with radical basis function kernel.
The best results has been obtained for $m=0.5\pi, \sigma=0.1, (\sigma_w=1)$ and circuit layers number $L=14$  set to twice the number of qubits.

\begin{table}
	\centering
	\caption{The best scores obtained with the considered methods.
		As the reference methods we consider logistic regression and
		supported vector machine with radical basis function kernel (SVM+RBF).
		The results of the reference methods come from~\cite{wilson2019quantum}.}
	
\begin{tabular}{cc}\hline \hline
	Method & Accuracy  \\ \hline 
	Logistic regression & .959 \\
	SVM+RBF & $\approx.99$ \\ 
	Quantum Generative Model Kernel & .9909 \\ \hline
\end{tabular}
\end{table}

\begin{figure}[h]
\centering
\includegraphics[]{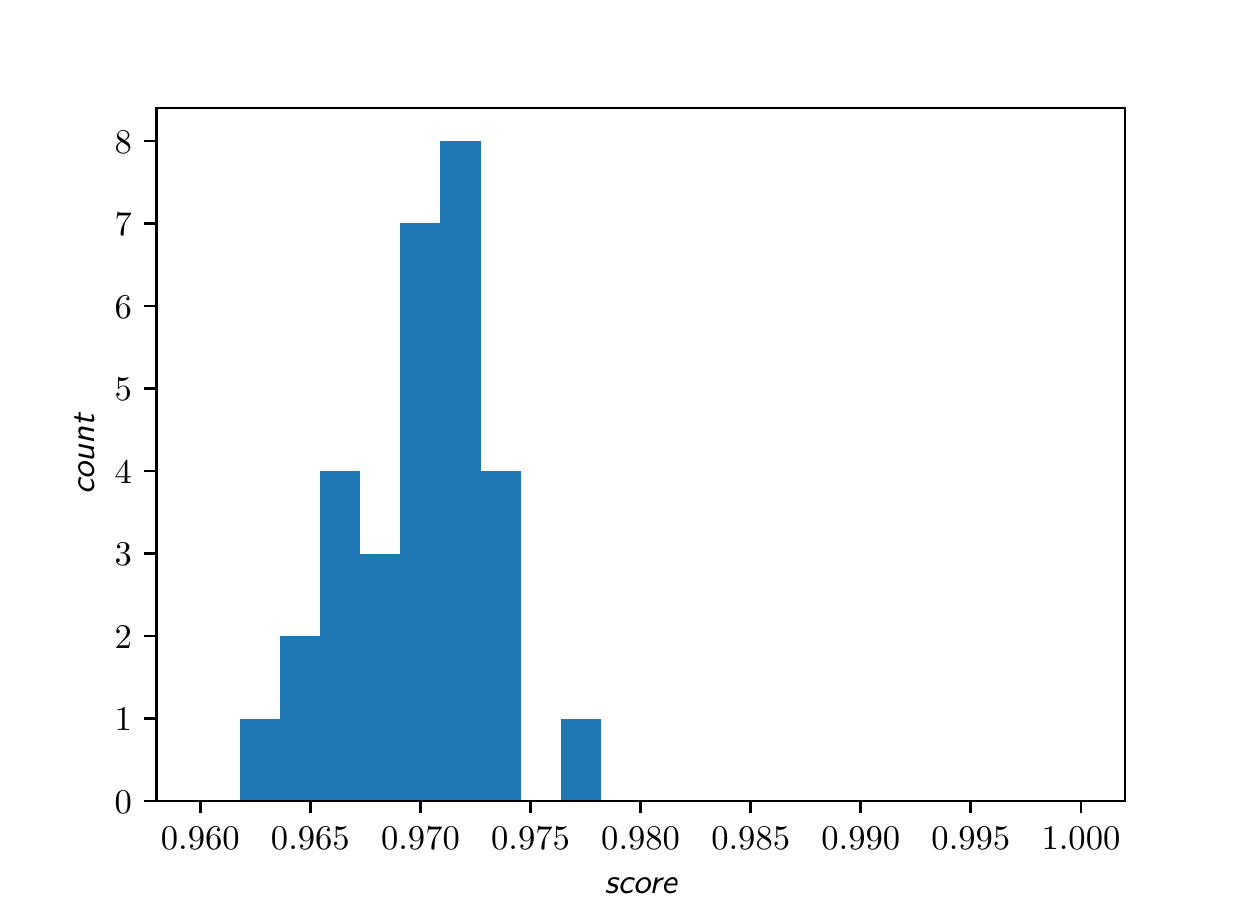}
\includegraphics[]{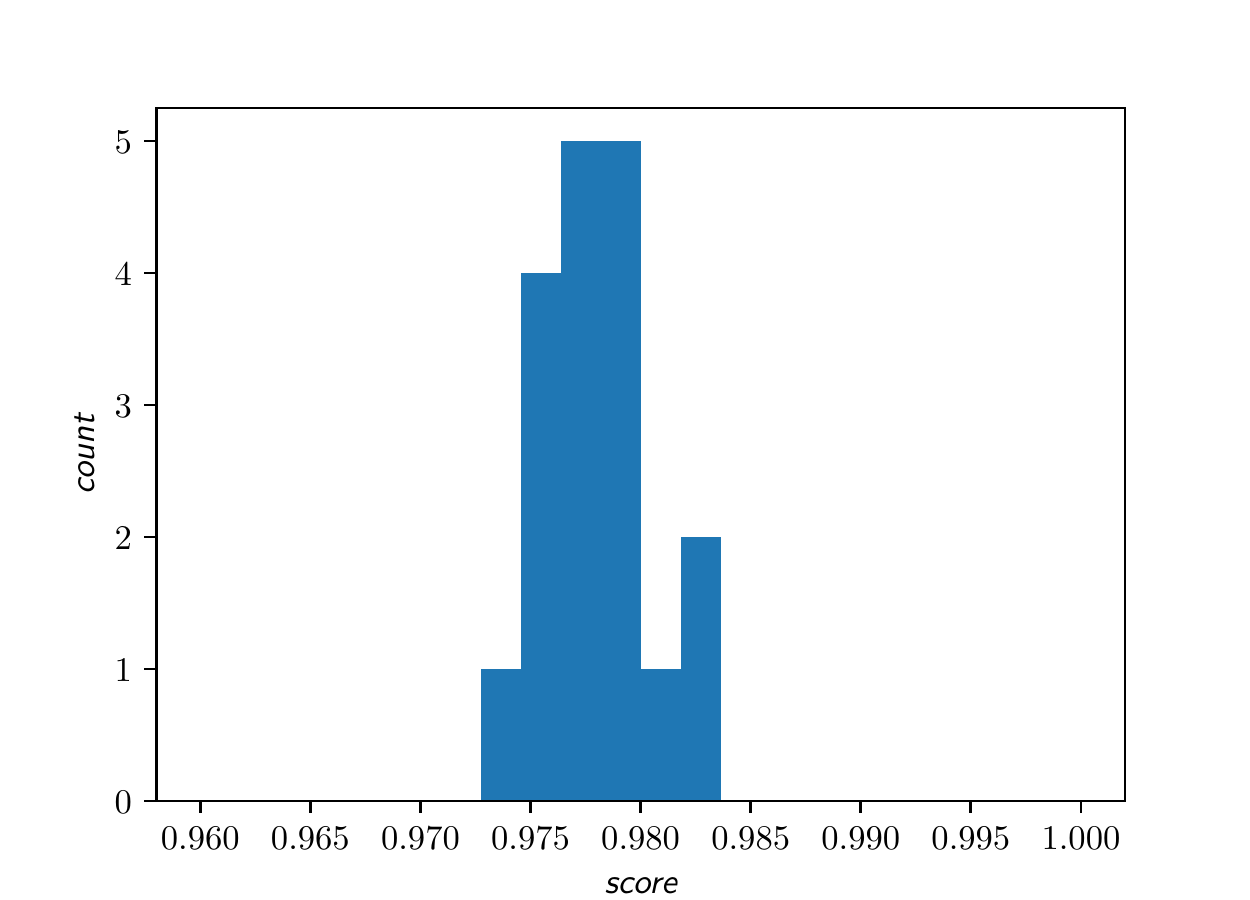}
\caption{Score histograms for number of random vectors $D$ equal to $500$ (top) and $1000$ (bottom). The scores aggregate various hyper-parameters.}\label{fig:hist1}
\end{figure}

\begin{figure}[h]
\includegraphics[]{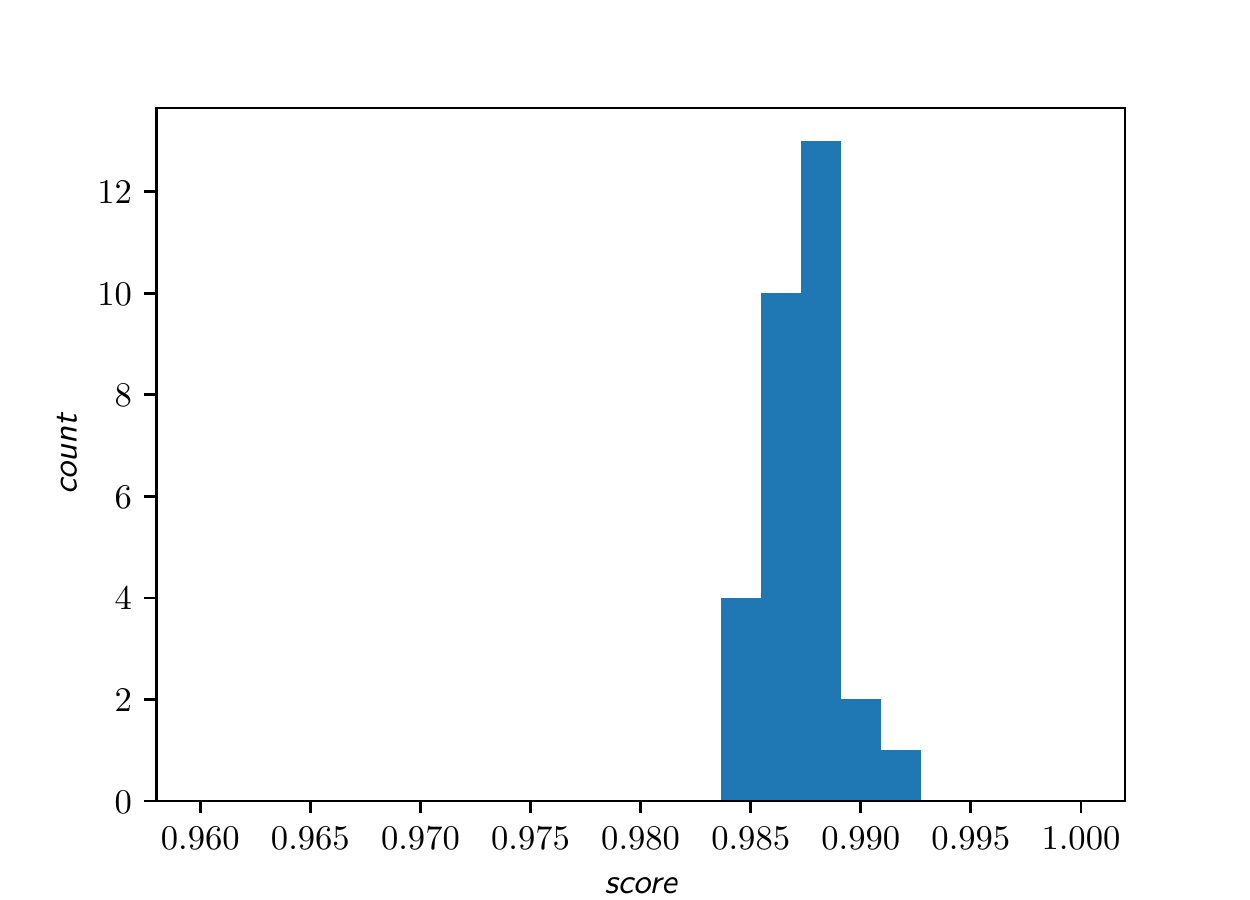}
\includegraphics[]{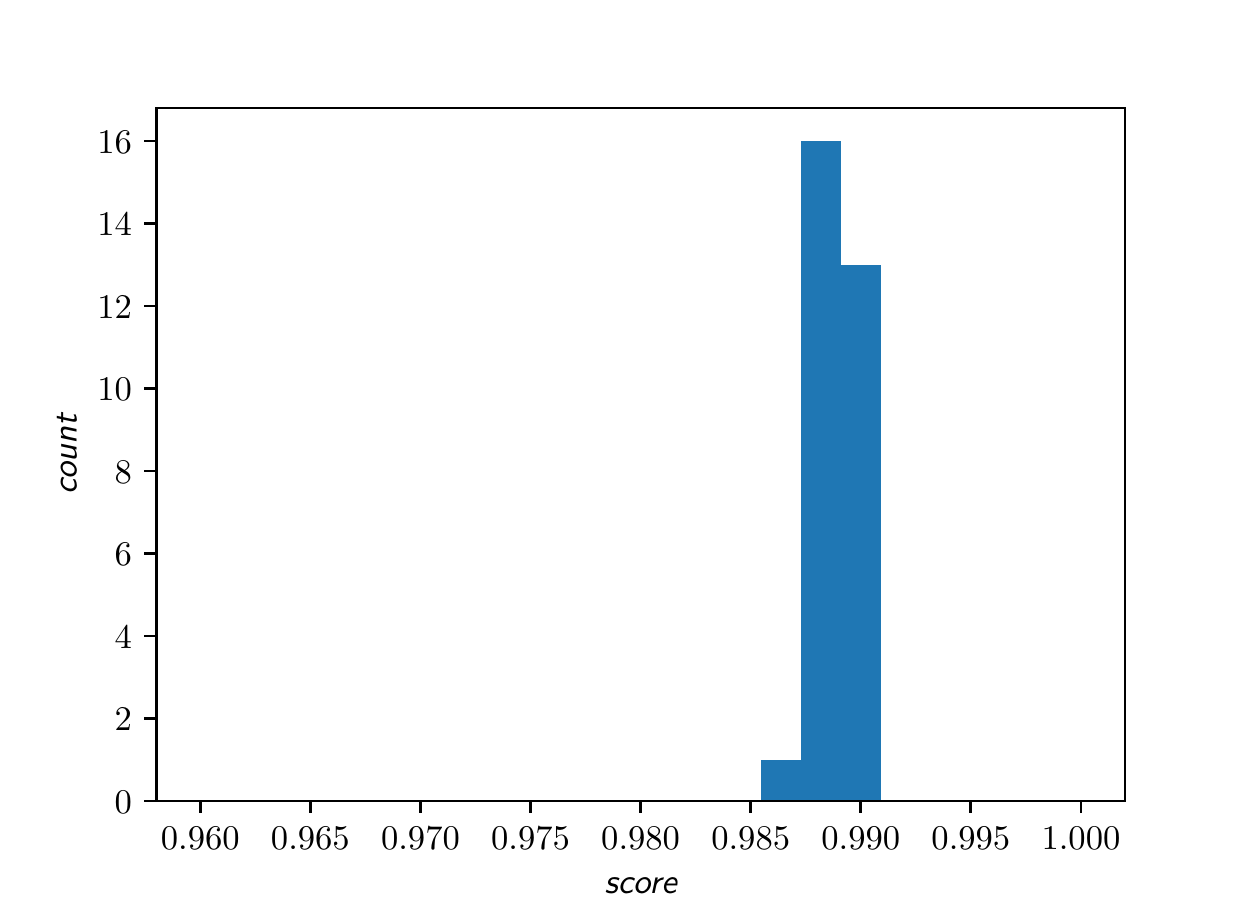}
\caption{Score histograms for number of random vectors $D$ equal to 8000 (top), 16000 (bottom). The scores aggregate various hyper-parameters.}\label{fig:hist2}
\end{figure}

\begin{figure}[h]
	\centering
	\includegraphics[]{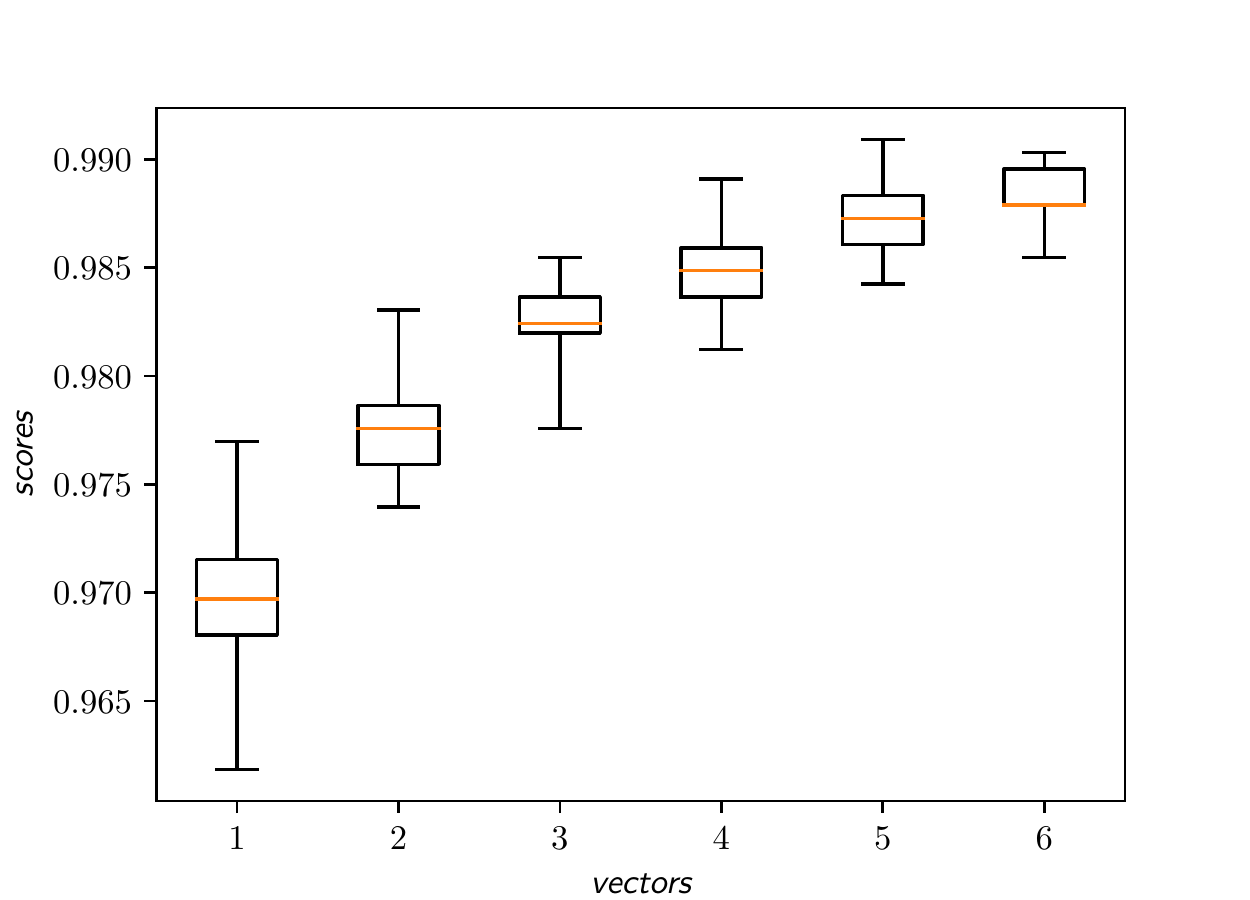}
	\caption{Scores obtained for number of random vectors $D$ equal to 500, 1000, 2000, 4000, 8000 and 16000.
		Whiskers reflect the min/max values.
		The highest obtained value is .9909.}\label{fig:scores}
\end{figure}

The histograms of the scores are presented in Fig.~\ref{fig:hist1} and Fig.~\ref{fig:hist2}.
The relation between best score and the number of random vectors $D$ is presented in Fig.~\ref{fig:scores}.


\section{Discussion}
There are three key points that we want to discuss, concerning simulation complexity, applicability to quantum data and the obtained scores.

Firstly,
the proposed method provides a link between quantum circuit Ans\"atze and machine learning kernels.
Any family of quantum circuits gives a new kernel.
For small quantum circuits this gives a quantum inspired kernel creation method.
For large circuits, that we can expect to yield a distribution that is hard to simulate,
we obtain kernels that can be considered non-classical.
However, we want to stress that the fact that simulating the circuits is time consuming does
not mean that sampling from the resulting distribution is as well.
Many families of random circuits are known to converge with length to easily
sampled distributions, in particular $k$-designs~\cite{harrow2009random}.

Secondly,
the method works with data encoded in quantum states. Thus, is compatible
with input data that is intrinsically of quantum nature. 
This may be an important feature if quantum simulation on quantum devices become
common,
and methods that handle output data directly would be desirable.
In particular, the kernel can be joined with methods that operate on states with known
preparation scheme but without the need to obtain the representation in
computational basis, as in 
VQE~\cite{peruzzo2014variational}.
Also, these are the most probable scenarios to require circuit sizes that cannot
be simulated classically in reasonable time.

Lastly,
the exemplary Ans\"atze seems competitive compared to established classical
methods.
The comparison is far from being a good argument for arguing supremacy over
classical methods, but supports optimistic view.
The selected problem is best handled with methods that harness spacial relations
in the image~\cite{mnist}.
For fair comparison these relations should be included.

The generative model that we choose in this work is only an example.
Apart from the circuit model there are other quantum computational models.
A natural turn for future work would be to look at other possibilities.
Another example of computationally universal model 
that can generate probability distributions with specific features
would be a quantum walk~\cite{sadowski2016lively}.
One could also turn to a general description of a quantum system given by
Schroedinger/Lindblad equation~\cite{pawela2016various}.
These models could yield different probability distributions, and thus be the
source of different kernels.

\bibliographystyle{unsrt}
\bibliography{arvis_v1}

\end{document}